\documentclass[useAMS,usenatbib]{mn2e}
\newcommand{\aap}{A\&A}

\newcommand{\mnras}{MNRAS}
\newcommand{\apjl}{ApJL}
\newcommand{\apjs}{ApJS}
\newcommand{\apj}{ApJ}
\newcommand{\aj}{ApJ}

\newcommand{\pasp}{PASP}

\usepackage[english]{babel}
\usepackage{graphicx}
\usepackage{color}
\usepackage{natbib}
\usepackage{amssymb}
\usepackage{lscape}
\usepackage{longtable}
\title[SN 2014J models ]{Photometric and spectroscopic observations, and abundance tomography modelling of the type Ia supernova SN 2014J located in M82}
\author[C. Ashall]{C.Ashall$^{1}$\thanks{E-mail:
c.ashall@2013.ljmu.ac.uk}, P. Mazzali$^{1,2,3}$, D. Bersier$^{1}$, S. Hachinger$^{4}$$^,$$^{5}$, M.Phillips$^{6}$, S. Percival$^{1}$, 
\newline \and P. James$^{1}$, K. Maguire$^{7}$\\
$^{1}$Astrophysics Research Institute, Liverpool John Moores University, IC2, Liverpool Science Park, 146 Brownlow Hill, \\  Liverpool L3 5RF, UK\\
$^{2}$Istituto Nazionale di Astrofisica-OAPd, vicolo dell'Osservatorio 5, 35122 Padova, Italy\\
$^{3}$Max-Planck-Institut f\"ur Astrophysik, Karl-Schwarzschild-Str. 1, D-85748 Garching, Germany\\
$^{4}$Institut f\"ur Theoretische Physik und Astrophysik, Universit\"at W\"urzburg, Emil-Fischer-Str. 31, 97074 W\"urzburg, Germany \\
$^{5}$Institut f\"ur Mathematik, Universit\"at W\"urzburg, Emil-Fischer-Str. 30, 97074 W\"urzburg, Germany\\
$^{6}$Carnegie Observatories, Las Campanas Observatory, La Serena, Chile\\
$^{7}$European Southern Observatory, D-85748 Garching bei M\"unchen, Germany}

\begin{document}

\date{Accepted 2014 September 22.  Received 2014 September 8; in original form 2014 July 29}

\pagerange{\pageref{firstpage}--\pageref{lastpage}} \pubyear{2014}

\maketitle

\label{firstpage}

\begin{abstract}
Spectroscopic and photometric observations of the nearby Type Ia Supernova (SN Ia) SN 2014J are presented.  Spectroscopic observations were taken -8 to +10 d relative to $B$-band maximum, using FRODOSpec, a multi-purpose integral-field unit spectrograph. The observations range from 3900 \AA\ to 9000 \AA. SN 2014J is located in M82 which makes it the closest SN Ia studied in at least the last 28 years. It is a spectrosopically normal SN Ia with high velocity features. We model the spectra of SN 2014J with a Monte Carlo (MC) radiative transfer code, using the abundance tomography technique. SN 2014J is highly reddened, with a host galaxy extinction of ${E(B-V)}$=1.2 (\textit{R}$_{V}$=1.38). It has a $\Delta$\textit{m}$_{15}(B)$ of 1.08$\pm$0.03 when corrected for extinction.  As SN 2014J is a normal SN Ia, the density structure of the classical W7 model was selected. The model and photometric luminosities are both consistent with $B$-band maximum occurring on JD 2456690.4$\pm$0.12.  The abundance of the SN 2014J behaves like other normal SN Ia, with significant amounts of silicon (12\% by mass) and sulphur (9\% by mass) at high velocities (12300 km s$^{-1}$) and the low-velocity ejecta (v$<$\,6500 km s$^{-1}$) consists almost entirely of   $^{56}$Ni.
\end{abstract}


\section{Introduction}

Supernovae are important and much-studied astrophysical events. For example they are the main producers of heavy elements in the universe, and type Ia supernovae  (SNe Ia) produce most of the iron-group materials \citep{FEelem}.  SNe Ia have also been confirmed as the best cosmological `standard candles' and as a result there has been a dramatic increase in the rate at which they are observed. They were integral in the discovery of the acceleration of the universe \citep{reiss98, perlmutter}, and are now an important cosmological probe in improving the understanding of the nature of the positive cosmological constant. However the true intrinsic properties of SN Ia are not yet fully understood, including their progenitor system and diversity in luminosity (e.g SN 1991bg, \citealt{1991bg}). There are currently two favoured progenitor scenarios. The first is a carbon/oxygen White Dwarf (WD) which accretes mass from a non-electron-degenerate companion star \citep{NomotoK97}. In this single degenerate (SD) scenario \citep{SD}, the WD can explode when it approaches the Chandrasekhar mass. There are several suggested ways in which this can occur, including a subsonic explosion (a deflagration) and a supersonic explosion (a detonation) as well as an explosion with a transition to a detonation. In the fully subsonic explosion there is not enough energy to fully power the SN Ia and in supersonic explosion there is too much $^{56}$Ni in the ejecta. Therefore the transition may be the correct explosion model for SN Ia \citep{DD}. Two further SD explosions models are fast deflagration and Sub-Chandrasekhar mass explosions \citep{w7, livne95}.  The other suggested progenitor model is a double degenerate (DD) scenario, where the SN results from the merger of two WDs \citep{WDWD}.  
\newline Thanks to the dramatic increase in observations of SN Ia, in the last two decades, it is now possible to obtain good spectral time series of them. These time series can span from before \textit{B}-band maximum to the nebular phase. One approach to fully understand the composition and progenitor system of an individual SN Ia, is to use MC radiative transfer code and the abundance tomography technique \citep{mazzali2000, Stehle2005}. This approach models early time observed spectra, by changing input parameters such as the chemical abundance, bolometric luminosity, photospheric velocity and time since explosion. The abundance tomography approach exploits the fact that with time deeper and deeper layers of the ejecta become visible. By modelling time series, spectral information about the abundances at different depths can be extracted, with this it is possible to  reconstruct the abundance stratification from the observational data. This approach directly links the theoretical models and observed spectra to help one get a true understanding of the early time evolution of a SN Ia. The MC radiative transfer code has been successfully used in modelling many SNe Ia including 2003du \citep{2003du}, 2004eo \citep{2004eo} and 2011fe \citep{mazzali2013}. 
\newline We present photometric and spectroscopic data taken with the Liverpool Telescope (LT) and Isaac Newton Telescope (INT) for SN 2014J, and then apply the aforementioned modelling techniques with the aim of inferring the ejecta properties on the SN. SN 2014J is a spectrosopically normal SN Ia, which has high velocity features. It is of particular interest as it is the closest SN Ia in at least the last 28 years, and possibly the closest in the last 410 years \citep{Foley}. As technology has dramatically improved in this time, this SN gives us a unique opportunity to intensely observe, analyse and model a SN Ia with modern technology. SN 2014J is located at RA=9:55:42 Dec=69:40:26.0 (J2000), and is in M82 which is at a distance of  3.77$\pm$0.66 Mpc. This value was obtained from the mean distance from NED\footnote{NASA/IPAC Extragalactic Database (NED) }, which used  2 methods PNLF (Planetary Nebula Luminosity Function) and TRGB (Tip of the Red Giant Branch) to derive it. However, it should be noted that \citet{Foley} derive a distance of 3.3 Mpc.  M82 is known for having a large amount of star formation; hence it has a large amount of dust \citep{M82}. Because of this, SN 2014J is highly and unusually reddened. It does not follow the average galactic reddening law of \textit{R}$_{V}$=3.1. Detailed investigations into the host galaxy extinction were produced by \citet{amanullah} and \citet{Foley}. With our modelling approach, we optimise the published extinction values which produce the best-fits of SN 2014J from the spectra.
\newline The paper starts with a report of the observations, which includes 12 spectra taken from --8d to +10d. In the following section (Section 3) we discuss the aperture photometry, including the SDSS \textit{g}$^{\prime}$\textit{r}$^{\prime}$\textit{i}$^{\prime}$ light curves. In Section 4 we present the fully reduced and calibrated spectra. The next section (Section 5) discusses the Monte Carlo (MC) radiative transfer technique. In Section 6 the modelled spectra of 10 early time epochs are presented and discussed. Afterwards (Section 7), we discuss the abundance stratification we have inferred. Finally , the results are summarised and conclusions are drawn from them (Section 8).

\section[Observations]{Observations}
SN 2014J was discovered on 2014/01/21.810 by S.J. Fossey at the UCL observatory \citep{14Jdis}. The Liverpool Telescope (LT) carried out detailed spectroscopic and photometric observations, starting from 2014/01/22. The LT is a 2.0 metre fully robotic telescope located at Observatorio del Roque de los Muchachos (ORM) on La Palma. Photometric observations were obtained using IO:O, an optical imaging camera which has a field of view (FOV) of 10 arcmin$^{2}$. The photometric images were acquired in three pass-bands (SDSS \textit{g}$^{\prime}$\textit{r}$^{\prime}$\textit{i}$^{\prime}$). Spectra were obtained using  FRODOSpec, a multipurpose integral-field unit spectrograph, at 10 epochs from --8d to +10d relative to B-band maximum. FRODOSpec consists of  blue and red arms which cover 3900-5700 \AA\  and 5800-9400 \AA\ respectively. 
 \newline The IO:O pipeline carries out basic reduction of photometric data; this consists of bias subtraction, trimming of the overscan regions and flat fielding. FRODOSpec has two independent reduction pipelines. The first one, L1, performs bias subtraction, overscan trimming and CCD flat fielding; whereas the second one, L2, is specific to FRODOSpec \citep{FRODOSpec}. It produces sky subtracted row stacked spectra, which were used in this paper. No host galaxy subtraction has been preformed when analysing the spectra, as no images were available. However, the sky subtraction routine in the FRODOSpec pipeline removes most of the flux from the M82, meaning that any flux which has not been subtracted should be negligible.      
\newline We also have two spectra obtained using the 2.5m Isaac Newton Telescope (INT), located at ORM on La Palma. The Intermediate Dispersion Spectrograph (IDS), a long-slit spectrograph on the INT, was used with the R1200Y grating and RED+2 camera. The INT observations were made with the slit at the parallactic angle. Table 1 is a log of spectroscopic observations of SN 2014J.

\begin{table}
 \centering
 \begin{minipage}{140mm}
  \begin{tabular}{ccccc}
  \hline
   Epoch & {MJD \footnote{Observation in MJD date}} & {Phase\footnote{Relative to B-band maximum}} & Exp time &Instrument\\
   &&(d)&(s)&\\
   \hline
2014-01-26 & 56684.12 & --8 & 500 &FRODOSpec\\
2014-01-27 & 56685.12 & --7 & 500&FRODOSpec  \\
2014-02-03 & 56692.01 & +0  & 600&FRODOSpec\\
2014-02-04 & 56692.90 & +1  & 600&FRODOSpec\\
2014-02-04 & 56693.13 & +1& 3$\times$120  &INT\\
2014-02-05 & 56693.89 & +2  & 600&FRODOSpec\\
2014-02-06 & 56694.96 & +3  & 600&FRODOSpec\\
2014-02-07 & 56695.89& +4 & 500 &FRODOSpec\\
2014-02-07 &56696.13 & +4 & 3$\times$120   &INT\\
2014-02-08&56696.92&+5&500&FROSOspec\\
2014-02-09 &56697.97 & +6  & 500&FRODOSpec\\
2014-02-11 & 56699.88& +8  & 500&FRODOSpec\\
2014-02-13 & 56701.877 & +10  & 500&FRODOSpec\\
\hline
\end{tabular}
 \caption{Log of spectroscopic observations of SN 2014J.}
\end{minipage}
\end{table}

\begin{figure}
\centering
\includegraphics[scale=0.4]{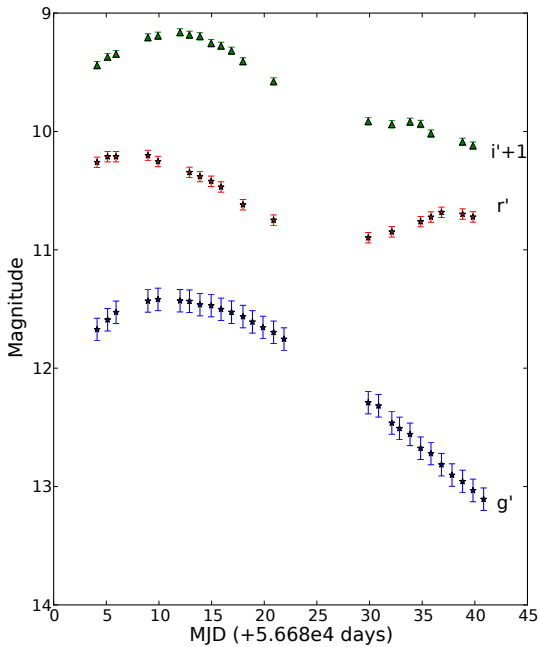}
\caption{Aperture photometry light curves, produced using LT data. SDSS \textit{g}$^{\prime}$\textit{r}$^{\prime}$\textit{i}$^{\prime}$ light curves are presented. The errors on the photometry appear to be constant; this is due to the catalogue error on the zero point star dominating.}
\label{fig:LC}
\end{figure}

\section{Photometry}

\begin {figure}
\centering
\includegraphics[scale=0.3]{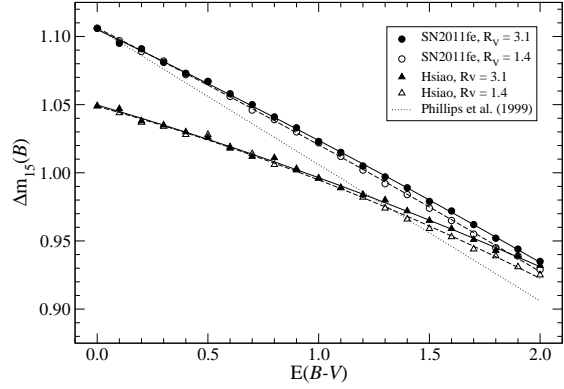}
\caption{Dependence of $\Delta$m(B)$_{15}$ on ${E(B-V)}$ for different values of \textit{R}$_{V}$, using SN 2011fe spectrophotometry and the Hsiao template.}
\label{fig:phillips}
\end{figure}

Photometric reduction was carried out using the \textsc{iraf}\footnote{\textsc{iraf} is distributed by the National Optical Astronomy Observatories, which are operated by the Association of Universities for Research in Astronomy, Inc., under cooperative agreement with the National Science Foundation.} package \textsc{daophot}. Instrumental magnitudes of the SN and stars within the field were obtained. In order to produce the calibrated magnitudes of the supernova, the colour terms and zero points were required. For the \textit{g}$^{\prime}$ \textit{r}$^{\prime}$ and \textit{i}$^{\prime}$ filters, the colour terms were obtained by using the standard star images, taken on the same night as the observations, and comparing their instrumental magnitudes to the APASS catalogue magnitudes\footnote{http://www.aavso.org/apass}. IO:O typically requires an exposure time of over 10 s for good photometry, due to the amount of time it takes for the shutter to open and close. We typically have an exposure time of 2 s in our photometric images. Therefore stars at the edge of the FOV will be exposed for a shorter period of time than ones in the centre of the field. To overcome this,  the zero points were calculated using a single star, RA=9:55:35 Dec=69:38:55 (J2000), close to SN 2014J.  The main source of error in the photometry was calculating the zero points; if the calibration star was saturated the magnitude was ignored. As we do not have pre-explosion images of M82 using the LT, we have not subtracted the host galaxy from the photometry. To calculate how much this host galaxy flux affects our photometry we used SDSS images of M82. We carried out aperture photometry, using the same aperture size as the LT images, on the location of the SN pre-explosion and a standard star in the FOV. From this we found that the highest value the photometry can deviate by, owing to the missing host galaxy subtraction, is 2.35\% mag in the \textit{i}$^{\prime}$-band, 1\% mag in the \textit{r}$^{\prime}$-band and 0.5\% mag in the \textit{g}$^{\prime}$-band. 
\newline SN 2014J was observed at a daily cadence from MJD 56680.20 to 56743.84.  \textit{g}$^{\prime}$\textit{r}$^{\prime}$\textit{i}$^{\prime}$ light curves were produced (Figure \ref{fig:LC}). There was a gap in observations between 13$^{th}$ Feb 2014 and 21$^{st}$ Feb 2014, due to poor weather conditions on La Palma.  \textit{g}$^{\prime}$-band maximum occurred on 03-Feb, which is consistent with the maximum in the \textit{B} pass-bands \citep{14JLC}. We found that \textit{g}$^{\prime}$-band maximum was 11.42$\pm$0.09 mag, \textit{r}$^{\prime}$-band maximum was 10.16$\pm$0.03 mag and \textit{i}$^{\prime}$-band maximum was 10.20$\pm$0.05 mag. The distance modulus we use for M82 was taken from NED, $\mu$=27.86 mag. Therefore, at maximum, SN 2014J had an absolute  \textit{g}$^{\prime}$-band magnitude of -16.44 mag, an \textit{r}$^{\prime}$-band magnitude of -17.68 mag and an \textit{i}$^{\prime}$-band magnitude of -17.66 mag, before correction for extinction. 
\newline The foreground Galactic extinction of SN 2014J is ${E(B-V)}$=0.05 mag (\textit{R}$_{V}$=3.1) \citep{amanullah,Foley}. A large host galaxy reddening of SN 2014J was inferred by \citet{amanullah} and \citet{Foley}. The values obtained for the colour excess using the CCM law \citep{CCMred} by \citet{Foley} were ${E(B-V)}$=1.24$\pm$0.1 mag and \textit{R}$_{V}$=1.44$\pm$0.06. However, \citet{Foley} state that the best solution to the extinction of SN 2014J is using a two-component circumstellar scattering and dust reddening model.  \citet{amanullah} use photometric comparisons to SN 2011fe to obtain the best-fit for the reddening values. Using the FTZ reddening law \citep{FTZ} between --5 to +35\,d relative to B-band maximum it was found that ${E(B-V)}$=1.29$\pm$0.02 mag and an \textit{R}$_{V}$ of 1.3$\pm$0.1 with a $\chi^2/dof=3.3$. It should be noted that the \textit{R}$_{V}$ value is much lower than the typical Galactic average of \textit{R}$_{V}$=3.1. We use a host galaxy extinction of ${E(B-V)}$=1.2 with \textit{R}$_{V}$=1.38, details of why we use these values can be found in Section 6. The attenuation of the flux in $B$ and $V$ is similar with the values of \citet{amanullah} and \citet{Foley}. 
\begin {figure}
\centering
\includegraphics[scale=0.4]{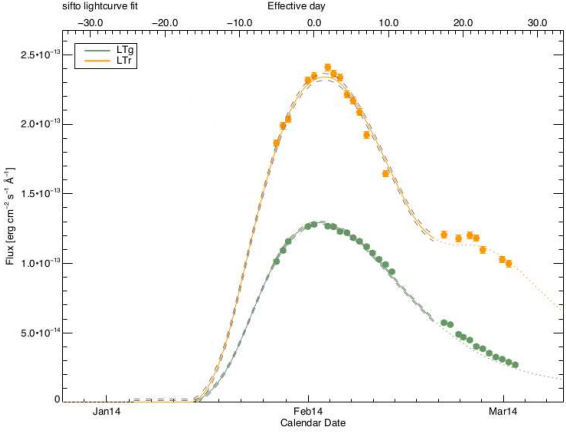}
\caption{SIFTO SN 2014J light curve fits}
\label{fig:sifto}
\end{figure}
\newline \citet{Foley} and \citet{marion} have published photometric and spectroscopic observations of SN 2014J. They report $\Delta m_{15}(B)$ to be 1.01-1.08 mag when corrected for host galaxy extinction, and 1.11$\pm$0.02 mag when not corrected for host galaxy extinction, respectively. We analysed the LT SDSS \textit{g}$^{\prime}$ and  \textit{r}$^{\prime}$-band photometry with the SiFTO \citep{sifto} light curve fitter to obtain the stretch, \textit{V}-band maximum and t$_{Bmax}$, see Figure \ref{fig:sifto}. Using the  \textit{g}$^{\prime}$ and  \textit{r}$^{\prime}$ bands we obtained a stretch of 1.083$\pm$0.06, which corresponds to a $\Delta$$m_{15}(B)$=0.88$\pm$0.08 using the relation from \citet{sifto}. However using only the  \textit{r}$^{\prime}$ band light curve produces a stretch of 1.035$\pm$0.08 and therefore a  $\Delta$$m_{15}(B)$ of 0.95$\pm$0.12. When corrected for host galaxy extinction, ${E(B-V)}$=1.2 (\textit{R}$_{V}$=1.38), using the relation from \citet{phillips1999}, the \textit{B}-band decline rate is found to be 1.00$\pm$0.06 or 1.07$\pm$0.08 for \textit{g}$^{\prime}$ and  \textit{r}$^{\prime}$ bands and \textit{r}$^{\prime}$ band respectively. These decline rates are consistent with those of \citet{Foley}.  However the correction of \citet{phillips1999} for obtaining ``reddening-free" $\Delta m_{15}$ values from the observed values was derived assuming an $R_{V}$\,$=$\,3.1. To check the sensitivity of this correction to \textit{R}$_{V}$, we carried out our own calculation of the effect of dust reddening on $\Delta$$m_{15}(B)$ for \textit{R}$_{V}$=3.1 and  \textit{R}$_{V}$=1.4 using the published optical spectrophotometry of SN 2011fe \citep{pereira} and the Hsiao SN Ia spectral template \citep{Hsiao}. The spectra were reddened for values of ${E(B-V)}$=0.0-2.0 using the \textsc{iraf} \textsc{deredden} task, which implements the CCM law. Synthetic magnitudes were calculated using the \citet{Bessell} \textit{B} passband,  and the \textit{B}-band decline rate was measured for each value of ${E(B-V)}$.  Figure \ref{fig:phillips} shows the results, including a comparison with the approximate relation given by  \citet{phillips1999}.  As is seen, the effect on $\Delta$$m_{15}(B)$ of changing the value of \textit{R}$_{V}$ from 3.1 to 1.4 is small.  For ${E(B-V)}$=1.2 and \textit{R}$_{V}$=1.38 application of these calculations to our SIFTO-measured decline rates gives $\Delta$$m_{15}(B)$=0.98$\pm$0.08 or 1.05$\pm$0.12,  for the  \textit{g}$^{\prime}$ and  \textit{r}$^{\prime}$ bands  and \textit{r}$^{\prime}$  band respectively, using the 2011fe spectra; and $\Delta$$m_{15}(B)$=0.95$\pm$0.08 or 1.02$\pm$0.12 using the Hsiao template. The value of the decline rate may increase when the full host galaxy subtraction can be carried out on the photometry. \citet{Foley} state that the \textit{V}-band maximum was 10.61$\pm$0.05, whereas our fitting obtains this to be 10.66$\pm$0.02. We obtain a \textit{t}$_{\textsc{b}max}$ which is consistent with the values found by \citet{Foley}.

\begin{figure}
\centering
\includegraphics[scale=0.75]{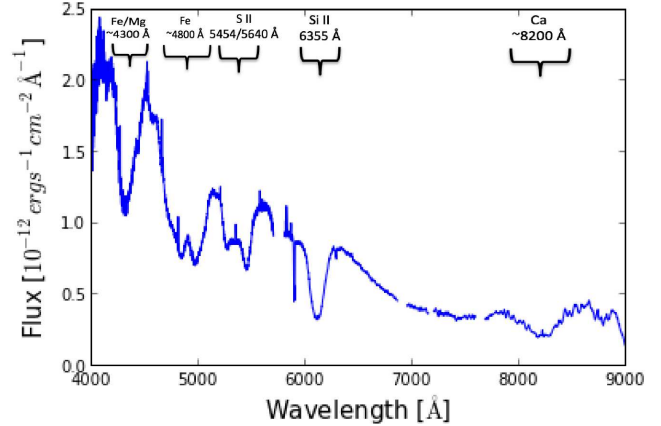}
\caption{An overview of the features of SN 2014J. This spectrum has been dereddened and was taken at B-band maximum.} 
\label{fig:lineplot}
\end{figure}

\section{Spectroscopy}
\begin{figure*}
\centering
\includegraphics[scale=0.45]{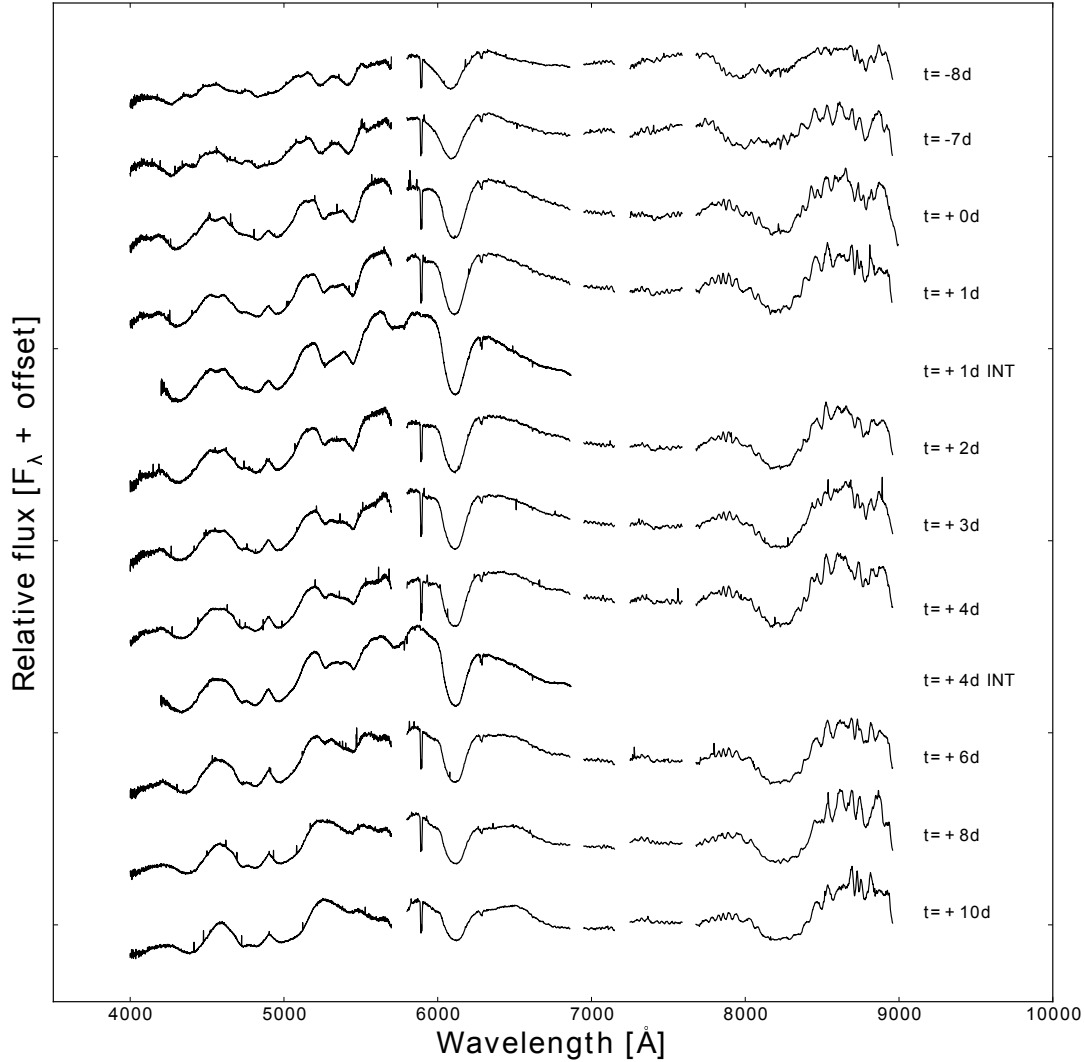}
\caption{All spectral observations of SN 2014J, LT and INT. The time is given relative to rest-frame $B$-band maximum. The spectra have not been corrected for reddening. There were no data collected between 5700-5800 \AA\  for the LT observations and the two atmospheric absorption lines have been removed. All of the plots have been offset by an arbitrary amount for the purpose of presentation.} 
\label{fig:allspec}
\end{figure*}

The LT data spectroscopic reduction was done in two halves, corresponding to the blue and red arms of FRODOSpec. Each spectrum was manually searched through to select fibres which had signal. An appropriate top threshold was applied, to ensure cosmic rays were not affecting the spectrum. The signals from these fibres were combined; the spectra were formed using the \textsc{onedspec} \textsc{iraf} package. The SN spectra were calibrated in flux using spectra of Feige34. The accuracy of the flux calibration process was confirmed by a successful calibration of the star back onto itself. This was done by running the calibration process on the observations of the standard star, and checking this against the \textsc{iraf} data for the star. The INT data were reduced using the \textsc{starlink} software \citep{starlink}. The spectrophotometric standard used to reduce the INT data was Feige66.  
\newline A full plot of a SN 2014J spectrum with the main absorption lines labelled is found in Figure \ref{fig:lineplot} and the early time spectral evolution of SN 2014J is plotted in Figure \ref{fig:allspec}. The spectra have a very strong Si II  6355 \AA\ absorption line. They also have high velocity features at early epochs; the main high velocity feature is exhibited by the Ca II IR triplet around $\sim$7900 \AA. Furthermore, at even earlier times up to --12d  \citet{Goobar} have found a strong high velocity feature of Si II 6355 \AA. Figure \ref{fig:allspec} demonstrates the extent of the reddening, which means that the flux in the blue arm is extremely suppressed. The evolution of the Si II 6355 \AA \ line can be seen in Table 2. As expected, as the photosphere recedes, the velocity of the ejecta decreases and hence the wavelength of the absorption line increases, although this may not always be apparent over smaller intervals. This is due to the resolution of the spectra and the errors on the velocity. The velocity gradient from the Si II 6355 \AA\ line between +0 to +9d relative to \textit{B}-band maximum is 58.8 km s$^{-1}$ day$^{-1}$. This makes SN 2014J a Low Velocity Gradient (LVG) SN Ia \citep{benetti2005}. It can be seen in the LT post maximum spectra at +6, +8 and +10d that P-Cygni re-emission redwards of the Si II 6355 \AA\ absoption has a flat top, which is also visible in the HST data at +11.3 d \citep{Foley}.  
\newline By using the correlation between the ratio of the depth of the silicon lines, 5972 and 6355 \AA, and absolute magnitude and therefore $\Delta$\textit{m}$_{15}(B)$ \citep{Nugent05} we have obtained a photometry-independent approximation for the $\Delta$\textit{m}$_{15}(B)$ of SN 2014J. We used the INT spectra from +1d to do this, as the INT data cover the 5972 \AA\ feature. We derive $\Delta$\textit{m}$_{15}(B)$=1.11$\pm$0.02, which is consistent with the values found by \citet{marion} and \citet{Foley}. Furthermore it is possible to use the EW of the Si II 5972 \AA\ to estimate the  $\Delta$\textit{m}$_{15}(B)$ \citep{hanchlingerEW}. Using this method  $\Delta$\textit{m}$_{15}(B)$ was found to be 1.27$\pm$0.15. 

\begin{table}
 \centering
 \begin{minipage}{140mm}
  \begin{tabular}{cccc}
  \hline
   Epoch & {MJD\footnote{Modified Julian Day number}} & {Phase\footnote{relative to $B$-band maximum (days)}} &{velocity\footnote{velocity of the Si 6355\AA\ absorption line}}\\
   &(d)&(d)&(kms$^{-1}$)\\
   \hline
2014-01-26 & 56684.12 & --8 & 12973 \\
2014-01-27 & 56685.12 & --7 & 12643\\
2014-02-03 & 56692.01 & +0  & 11758\\
2014-02-04 & 56692.9 & +1  & 11697\\
2014-02-05 & 56693.89 & +2  & 11686\\
2014-02-06 & 56694.96 & +3  & 11552\\
2014-02-07 & 56695.89& +4 & 11640\\
2014-02-08 & 56696.89 & +5  & 11412\\
2014-02-09 &56697.97 & +6  & 11529\\
2014-02-11 & 56709.87 & +8  & 11408\\
2014-02-13 &  56701.877 & +10  & 11228\\
\hline
\end{tabular}
 \caption{ Log of the Si 6355\AA\ absorption line velocities}
\end{minipage}
\end{table}

\section{modelling technique}
By observing spectra at frequent intervals it is possible to study the detailed properties of SNe as a function of depth, as the photosphere recedes into the material with time and more and more of the ejecta mass becomes visible. We express this by a photospheric velocity decreasing in time, as velocity can be used as a co-moving spatial coordinate. This is because high velocities correspond to large radii, due to the homologous expansion of SNe Ia $\sim$10 s after the explosion. This can be approximated by Equation \ref{eq:radii}, where r is the distance from the centre of the explosion, v$_{ph}$ is the velocity and t$_{exp}$ is the time from explosion. 

\begin{equation} 
r={v_{ph}}\times{t_{exp}}
\label{eq:radii}
\end{equation}

The early time spectra are of particular interest as they change  significantly over very small epochs, allowing us to infer abundance information about the outermost layers of the explosion. Carrying on with the modelling up to post-maximum phases, the abundance stratification is inferred for most of the ejecta except for the core layers, which become visible only in the nebular, late phase. This phase must be modelled separately (which is beyond the scope of this paper), owing to the non-local-thermodynamic-equlibrium state of the plasma at late epochs.
\newline A useful way to model early time spectra, which still obtains information from observational data, is to use MC radiative transfer codes. Using this approach over a number of high and frequent epochs leads to an accurate abundance distribution of the SN. The MC radiative transfer code here is based on code originally written by \citet{Abbott1984} in relation to stellar winds. This was adapted by \citet{mazzalilucy93} for early time SN spectra, and further improved by \citet{Lucy1998} and then \citet{mazzali2000}. 
\newline We obtain optimally fitting synthetic spectra for the photospheric phase as \citet{Stehle2005} did with their abundance tomography approach. This technique has been used to model a number of SNe, including SN 2002bo \citep{Stehle2005}, SN 2010jn \citep{hachinger2013} and SN 2011fe \citep{mazzali2013}.   
\newline The code assumes 1D spherically symmetric ejecta and a grey photosphere from which radiation is emitted, and calculates the interaction of photons within the expanding SN ejecta. The code simulates the propagation of emitted photons by considering them as photon packets, which can undergo Thomson scattering and line absorption. At the photosphere, below which the ejecta are assumed to be optically thick, the outward flowing radiation is assumed to be from a blackbody. This assumption can cause excess flux in the IR and red side of the spectra at later epochs. However, it has the advantage that the abundances can be derived without the knowledge of radioactive heating below the photosphere. Furthermore, since all of the important absorption lines for the abundances are in the UV/optical this excess flux in the IR has little effect on the results.  The exact $^{56}$Ni distribution below the photosphere is therefore irrelevant to the calculation, it is only important that the bulk of the $^{56}$Ni of the SN is below the photosphere. Therefore the code can only be used up to $\sim$14 days past \textit{B}-band maximum, after this the assumptions of the code are weak.  
\newline The code has a selection of input parameters; the bolometric luminosity, ejecta velocity and chemical abundance stratification. The typical process of stratification tomography consists of modelling the earliest spectrum then moving inwards until the last photospheric spectrum is reached. The initial abundances are set by an educated approximation from a typical explosion model. The luminosity is then iterated until it matches the flux of the observed spectra. This is followed by the iteration of  the velocity to fit the observed blue-shifts of the P-Cygni features, which in turn fits the position of the lines. Finally the abundances are changed to ensure that the absorption line strengths are all modelled. The abundance for a new inner layer is calculated. The abundance tomography technique ensures that this spectrum shows absorption from the outer layers but also additional layers which are above the new photosphere and below the old one. This modelling process is repeated for each spectrum and previous abundances in outer layers may need to be changed in order to fit the later spectra, in which case the entire fitting process is repeated.  
\newline Choosing a reasonable density profile is important in producing a physically meaningful and well-fitting model. The density distribution chosen for SN 2014J is the W7 model \citep{w7}. This is a fast deflagration explosion of a Chandrasekhar mass C+O WD. The deflagration wave synthesizes 0.5-0.6 M$_{\sun}$ $^{56}$Ni in the inner layer of the star, which is enough to power the light curve of the supernova \citep{w7}. The W7 model was selected as SN 2004eo, SN 2003du and SN 2002bo can all be reasonably modelled with this density profile. Therefore we can see if there is continuity in results. A higher density can lead to enhanced absorption lines, therefore it is important to select the appropriate one for the explosion. The most marked effects of the density on photospheric spectra occur at the earliest epochs, when the density of the outermost layers strongly modulates e.g. the UV flux (e.g. \citealt{mazzali2013}). Therefore, moderate deviations in the outer density profile will not affect the results obtained from the regions explored in this paper, as we begin with the modelling at an epoch of -8 days relative to $B$-band maximum. Still, in follow-up papers we will test density profiles of double-degenerate and sub-Chandrasekhar-mass models.

\section {models}
Ten epochs of LT data have been modelled using the abundance tomography technique. All of the spectra have been dereddened using the CCM law. In this paper we use a host galaxy extinction of ${E(B-V)}$=1.2 mag (\textit{R}$_{V}$=1.38). These values are where our model produce the best fits, and are also consistent with the values of \citet{amanullah} and \citet{Foley}. The MC radiative transfer code is successful at modelling a variety of  SN, therefore the models of SN 2014J are a good indication that the derived reddening values are correct. One shell models were produced with different values of E(\textit{B}-\textit{V}), and it was found that if this value was increased the input luminosity of the model had to be increased. This meant there was too much flux in the model and it peaked in the UV rather than the optical. 
\newline The time since explosion is one of the input parameters needed for the modelling process, therefore the rise time of SN 2014J is needed. There are pre-discovery images of SN 2014J, in which it first appears at Jan. 14.75 UT \citep{zheng} which gives it a bolometric rise time of $\sim$20 days. The main input parameters can be found Table 3.  
\newline The main limitations of the analysis concern the high velocity outer layers of the ejecta, because we do not have early time spectra, the extinction values, the distance to SN 2014J and the lack of UV data. Although we may expect uncertainty to be in the order of $\sim$ 10\%, the biggest uncertainty is due to the lack of early UV data.  The UV data will be modelled in a follow up paper using the HST data.
\newline This section will analyse each epoch of the models and observations. In the models we parametrise the iron-group content in terms of two quantities, Fe and $^{56}$Ni at the time $t=0$. This gives us the abundance of Fe, $^{56}$Co and $^{56}$Ni at any point of time, assuming that the abundances of directly-synthesised Co and Ni are negligible. Therefore, it should be noted that any Fe discussed in this section is stable iron.   

\begin{table}
 \centering
 \begin{minipage}{140mm}
  \begin{tabular}{ccccc}
 \hline
 Epoch  \footnote{Relative to \textit{B}-band maximum} & Epoch \footnote{Days after the explosion} & Velocity & Bol. Lum&Temp\\
  t$_{\textit{B}}$& t$_{exp}$ & v$_{ph}$&& T$_{BB}$\\
 (d)&(d)&(km s$^{-1}$)&log(L$_{\sun}$)&(K)\\ 
 \hline
 -8 & 12 &12300 & 9.295 & 10200\\
  -7 &13 & 11990 & 9.408 & 10100\\
 +0 & 20 & 9480 & 9.455 & 9200\\
+1  &21 &8970 & 9.46&9000 \\
+2  & 22 &8440 & 9.43&8900\\
+3  & 23 & 7930 & 9.39&8800\\
+4  & 24 &7480& 9.355&8600\\
+6  & 26 & 6500 & 9.32& 8400\\
+8   & 28 &5450 & 9.250&8100\\
+10   & 30 & 4400 & 9.22&7800\\
\hline
\end{tabular}
\caption{Input parameters and calculated converged temperature.}
\end{minipage}
\end{table}

\begin{figure*}
\centering
\includegraphics[scale=0.8]{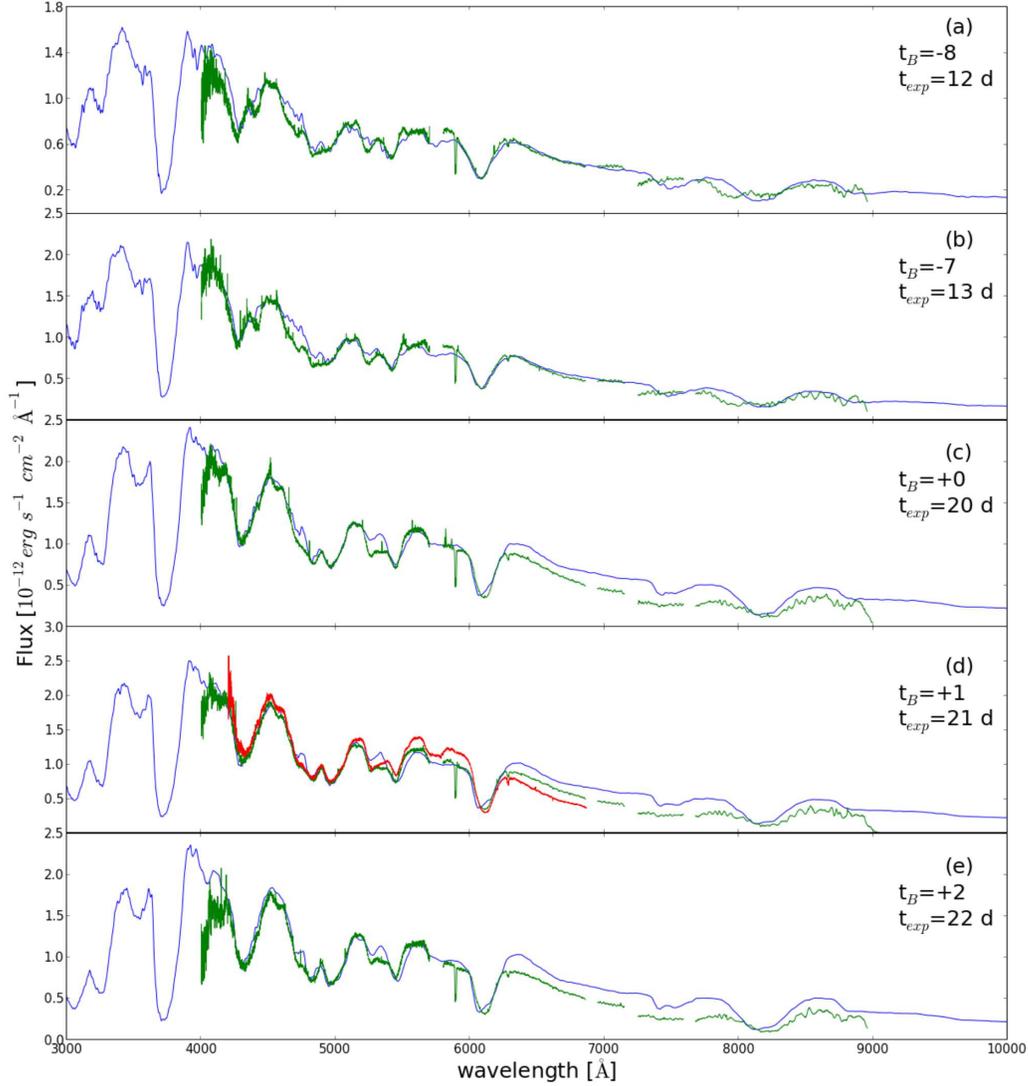}
\caption{Modelled (blue) and observed spectra (green) of SN 2014J. Plot (a) is from -4d, (b) is from -3d, (c) is +0d maximum, (d) is +1d and (e) is +2d. All of the dates are relative to B-band maximum. All spectra have been dereddened. The red spectrum at +1 d is from the INT.} 
\label{fig:model1}
\end{figure*}

\subsection{Day --8}
The first spectrum was observed 8 days before B-band maximum, t$_{exp}$= 12 days, see Figure \ref{fig:model1}. As this spectrum is before \textit{B}-band maximum it has a high photospheric velocity, v$_{ph}$=12300 km s$^{-1}$, the effective temperature is 10200 K and the bolometric luminosity is $10^{9.295}$ L$_{\sun}$. The strongest photospheric absorption lines, which we have indicated in Figure \ref{fig:lineplot} (Section 4), are dominant from the beginning of the time series to at least a week after maximum.
\newline There are strong Si II 6355 \AA,  S II 5454 \AA\ and 5640 \AA\  features. Due to this the model requires a photospheric abundance of 12\% Si and 9\% S by mass, with 13\% Mg to produce the Mg II  4481 \AA\  line. The Mg II  4481 \AA\  line is the dominant line in the 4300 \AA\ feature. The O abundance is 60\% and the Ca is 1\%. There are strong Ca II features in this modelled spectrum, including a large absorption line in the H\&K feature in the near UV (3934, 3968 \AA). This absorption line is produced by our models in a strength consistent with the HST data \citep{Foley}. Although we do not cover this UV Ca line in our observations, we do have the Ca II IR triplet from which we can infer the Ca abundances.  The photospheric Ca II IR triplet is modelled successfully, but there is no attempt to model the high velocity feature. An iron abundance of 4\% is required to produce the  Fe III  5150 \AA\ absorption line.  The $^{56}$Ni abundance is at 0\%, because the ejecta at this epoch are still in the high velocity outer layers of the photosphere.  Ti+Cr are at a photospheric abundance of 0.3\% by mass. There is a low abundance of C at this high velocity as there are no C spectral features in the optical data, suggesting that all the carbon may be in earlier time spectra, although \citet{marion} do find C I at 1.0693 $\mu$m. However, if we add C into the models we find this produces deeper and wider C absorption lines. Therefore the C found in the NIR could be due to a very small abundance. There is a small absorption line in the optical observed at $\sim$6200 \AA, where one would expect C, however this line is too narrow and not a C feature. The model at this epoch produces a particularly good match for the Fe II/Mg II 4200 \AA, S II 5640 \AA\  and Si II 6355 \AA\ features. 

\subsection{Day --7}
The second spectrum was taken --7d from \textit{B}-band maximum, t$_{exp}$= 13 days. The luminosity at this epoch is $10^{9.408}$ L$_{\sun}$ and  v$_{ph}$=11990 km s$^{-1}$. There is very little variation between the first two epochs as they are only taken 1 day apart. The main chemical changes from the previous epoch are that Si has increased to a photospheric mass abundance of 15\% and S to 10\%. The $^{56}$Ni has also increased to 3\%. The $\sim$4200 \AA\ feature is predominately Mg II 4481 \AA\ with smaller contributions from Fe III 4419 \AA. The Mg abundance is 10\% and the O abundance has decreased to 56\%. The Si III 4550 \AA\ line is successfully modelled, and this is the last FRODOSpec spectrum with a prominent Si III feature. 

\subsection{Day+0} The third spectrum was taken on the night of \textit{B}-band maximum, t$_{exp}$= 20 days, refer to Figure \ref{fig:model1}. The luminosity is $10^{9.455}$ L$_{\sun}$  and photospheric velocity v$_{ph}$=9480 km s$^{-1}$. The Si and S abundances have increased relative to the previous epoch to photospheric mass abundances of 17\% and 12\% respectively, and the $^{56}$Ni has stayed constant at 3\%.  $^{56}$Fe has also increased by 2\% to 6\%. Conversely, the Mg and O abundances have decreased to 5\% and 55\% respectively. There is now a notable excess in flux in the red side of the spectrum due to the black body approximation.  The \textit{B}-band modelled absolute unreddened magnitude is  --18.79. At this epoch the modelled Si III 4550 \AA\  absorption line is not as strong as the previous epochs. Furthermore, in the previous spectrum the 4800 \AA\ feature which have dominant Si II 5056 \AA\ and Fe II 5169 \AA\ absorption lines are merged into one, whereas at this epoch they have two distinct minima. Ca absorption is now more prominent in both the model and the observations, and is seen in the Ca II IR triplet at $\sim$8200 \AA\ .

\subsection{Day+1} The next spectrum in the series was observed on +1 d relative to \textit{B}-band maximum,  t$_{exp}$= 21 days. It has a  luminosity of $10^{9.460}$ L$_{\sun}$  and a photospheric velocity of v$_{ph}$=8970 km s$^{-1}$. This spectrum is at maximum bolometric luminosity, which is consistent with it being between \textit{B}-band and \textit{V}-band maximum. Although the INT and LT spectra differ slightly, predominantly in the red side of the spectrum, the main absorption lines in the blue side of the optical are similar between the spectra. Due to this the abundances obtained in our analysis would not change if we were to model just the INT or LT spectra.  The abundances we derive from this spectrum are very similar to the previous epoch; the Si decreases to 15\%, the S to 10\% the O to 57\%. $^{56}$Ni has increased  to 11\% and $^{56}$Fe is constant at 6\%. The effective temperature at this epoch is 9000 K, which is 200 K lower than the previous spectrum. The Mg abundance has decreased to 0\%. The S II  5454 \AA\ feature is not as strong in the model as the observed spectrum, however increasing S abundance would enhance the S II 5640 \AA\ line. There is excess strength of O I and Mg II  at 7773 \AA\ and 7896 \AA\ which could be an indication that there is excess mass at this velocity. This feature occurs in most of the epochs in the model, and is more dominant in late time spectra.   

\subsection{Day +2} The fifth spectrum was observed on t$_{exp}$= 22 d, see Figure \ref{fig:model1}.  Its luminosity is $10^{9.430}$ L$_{\sun}$  and photospheric velocity is v$_{ph}$=8440 km s$^{-1}$. The effective temperature is 8900 K. The abundances at this epoch are very similar to the previous one, except for $^{56}$Ni which begins to increase dramatically to 36\%, S which decreases to 4\%; Si and O also decrease  to 10\% and 43\% respectively. At this epoch Ca stays constant at 2\%. The $B$-band unreddened modelled absolute magnitude of this spectra is --18.85. At this epoch the  $\sim$4200 \AA\ feature is still dominated by the Fe III 4419 \AA\ and Mg II 4481 \AA\ lines. 
   
\begin{figure*}
\centering
\includegraphics[scale=0.8]{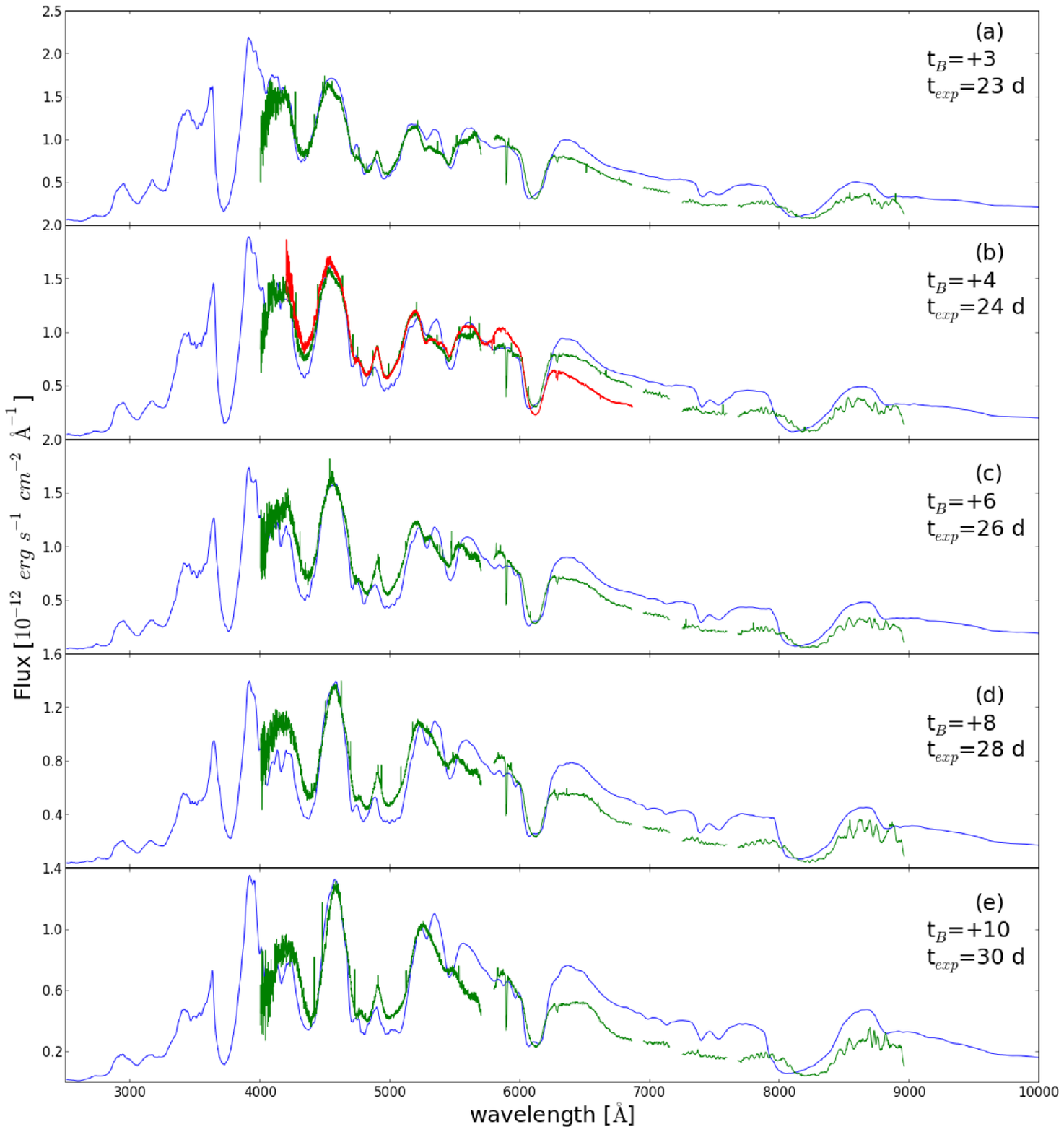}
\caption{Modelled (blue) and observed spectra (green) of SN 2014J. plot (a) is from +3d, (b) is from +4d, (c) is +5d, (d) is +6d and (e) is +8d. All of the dates are relative to B-band maximum. All spectra have been dereddened. The red spectum of +4 d is from the INT.} 
\label{fig:model2}
\end{figure*}

\subsection{Day +3}
The next spectrum was observed on t$_{exp}$= 23 days, refer to Figure \ref{fig:model2}. The luminosity at this epoch is $10^{9.390}$ L$_{\sun}$,  the photospheric velocity is v$_{ph}$=7930 km s$^{-1}$, the effective temperature is 7900 K and the \textit{B}-band modelled absolute magnitude is --18.622.  The modelled  S II 5640 \AA\ feature is stronger than the observed one; this is a problem which consistently occurs in the model. To make the S II the same strength as the observed one it would require the abundance of S to be reduced in the early-epoch models. However we have chosen to fit the early-time spectra rather than the late time ones, as these can lead us to more information about the high velocity abundances. At this epoch the O has a photospheric abundance of 0\%. The Si and S have also decreased to an abundance of 4\% and 2\% by mass respectively.   The Fe II 4420 \AA\ and Si II 6355 \AA\ lines are modelled successfully. The wide deep absorption line at 8200 \AA\  is the Ca II IR triplet absorption line, the calcium has been reduced to a 1\% by mass to decrease the strength of this line. At this epoch $^{56}$Ni begins to dominate and is at 88\%. Given the sudden jump in the $^{56}$Ni abundance we can give the constraint that the $^{56}$Ni distribution will extend to $\sim$8000 km s$^{-1}$. We thus predict that the characteristic line width of the iron emission in the nebula spectra will be in the order of $\sim$8000 km s$^{-1}$ \citep{mazzali98}.

\subsection{Day  +4}
The spectrum from +4 d has a luminosity of $10^{9.345}$ L$_{\sun}$ and  photospheric velocity of v$_{ph}$=7480 km s$^{-1}$. There is a similar discrepancy between the INT and LT data as in the +1 d plot, however once again this does not affect the abundances we obtain from our fits.  From this epoch the absorption lines are beginning to be stronger than the observed ones. For example, the modelled Fe/Mg 4300 \AA\ and Fe 4800\AA\ features, which are produced by dominant Fe III 4419 \AA\  and Fe II 5018 \AA\  absorption lines. However, to refine this fit requires the abundances of these lines at high velocity epochs to be reduced, affecting the early time spectra. Therefore, we suggest that the excess absorption could be due to too much mass at v$_{ph}$=7480 km s$^{-1}$.  Part of the excess in strength of the $^{56}$Fe lines could also be due to the decay of $^{56}$Ni. Due to this excess strength in Fe, its photospheric abundance is now at 0.1\%. The Si and S abundances have decreased to 0\%, and the $^{56}$Ni is at 99\%. The effective temperature at this epoch is 8600 K. 

\subsection{Day +6}
The eighth spectrum was taken 26 days after explosion. It has a luminosity of $10^{9.32}$ L$_{\sun}$, a photospheric velocity of v$_{ph}$=6500 km s$^{-1}$ and an effective temperature of 8400K. The $^{56}$Ni photospheric abundance is at 99\%. The Fe II 4340 \AA, S II 5606 \AA\  and Fe III 4419 \AA\  absorption lines are much deeper in the model than the observations. The unreddened modelled \textit{B} and \textit{V} magnitudes are 9.5 mag and 9.1 mag respectively.  

\subsection{Day +8}
The next spectrum has a luminosity of $10^{9.250}$ L$_{\sun}$  and  photospheric velocity is v$_{ph}$=5950 km s$^{-1}$. The $^{56}$Ni  abundance has stayed constant. The difference between the model and observation begins to differ even more, as shown by the Fe II 4549 \AA\ line in the 4800 \AA\ feature. At this epoch there is a significant amount of $^{56}$Ni above the photosphere. Therefore it is not unexpected that the difference between the models and observed spectra begins to increase. 
 
\subsection{Day +10}
The final epoch that was modelled is from +10 d from \textit{B}-band maximum. It has a luminosity of $10^{9.22}$ L$_{\sun}$, a  photospheric velocity of v$_{ph}$=5450 km s$^{-1}$ and an effective temperature of 7800 K. At this epoch there is no Si or S abundance, and the  $^{56}$Ni  abundance is the most dominant, at almost 100\%. The Ca abundance has now decreased to 0\%. 

\begin{figure*}
\centering
\includegraphics[scale=0.7]{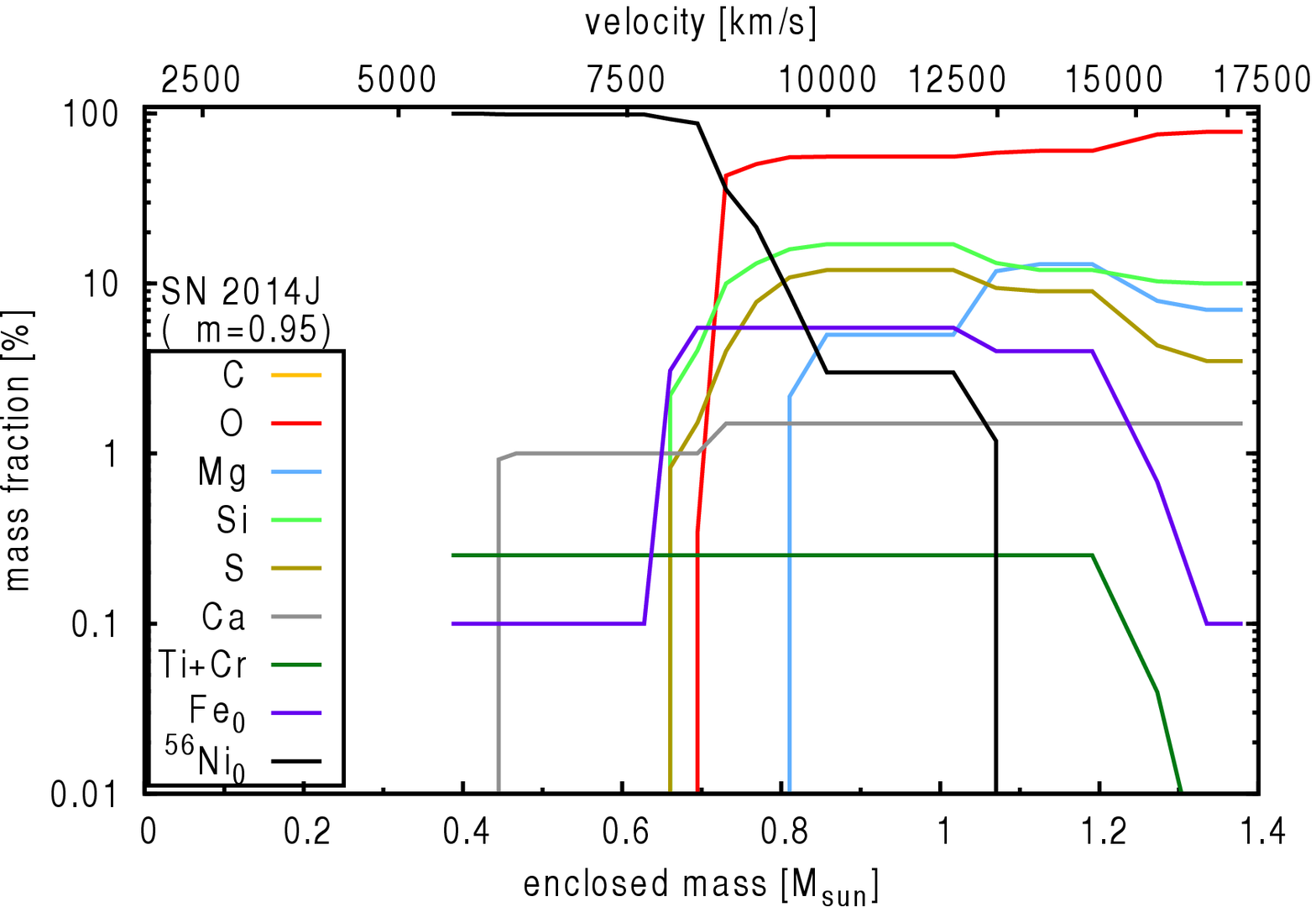}
\caption{The abundance distribution of SN 2014J. The Ni/Co/Fe abundances are given in terms of mass fractions of $^{56}$Ni ($^{56}$Ni$_{0}$) and stable Fe (Fe$_{0}$) at $t=0$. In the spectral models no stable Ni or Co and no radioactive Fe is assumed to be present.} 
\label{fig:abundance}
\end{figure*}

\section{abundance stratification}
We produce an abundance tomography distribution plot for the photospheric layers of the ejecta, see Figure \ref{fig:abundance}. This plot demonstrates how the abundances in the early times, from --8 d to +10 d, of SN 2014J develops. We cannot definitively confirm the abundances of the outermost layers of the ejecta as we do not have spectra before --8d. The abundances in the outer shell are slightly unusual, in that there is no carbon abundance. The other notable abundance is Fe which begins at 0.1\% and rises to 6\%. We have attempted to increase this, but doing so dramatically strengthens the Fe 4800 \AA\ feature at all epochs. Therefore the initial abundances of the outer shell, which has a velocity of 14800 km s$^{-1}$, are Si 10\%, O 78\%, S 4\%, $^{56}$Ni 0\%, Mg 7\%, C 0.0\%, $^{56}$Fe  0.1\%, with heavier elements making up the remaining abundance. 
\newline  At high velocities, between 8440-14800 km s$^{-1}$, there is a large oxygen abundance which starts at 78\%. The Si distribution starts at 10\% due to the strong Si II 6355 \AA\ feature, and it increases to 17\% at 9480 km s$^{-1}$ before decreases to 0\% at 7480 km s$^{-1}$. Sulphur also follows a similar distribution with respect to velocity, although it always has a smaller abundance than Si. The sulphur distribution starts at 3.5\% before peaking at 12\%. The basic abundance evolution of the ejecta involves O dominating followed by the IME and then by the heavy elements. In the abundance distribution plot, Figure \ref{fig:abundance}, the Fe starts at 0.1\% and rises to 6\%. In Figure \ref{fig:abundance} the IME elements are significant at high velocities. From this it can be inferred that the lighter elements may be at even higher velocities. Therefore earlier spectra are needed to gain information about these lighter elements. As expected from a normal SN Ia explosion $^{56}$Ni dominates the abundance. This happens between 8440-7930 km s$^{-1}$, where the  $^{56}$Ni goes from a photospheric mass fraction of 36\% to 83\%. The Ti+Cr abundances stay at a constant level throughout the whole explosion at 0.3\%. Calcium starts at 2\% at 12300 km s$^{-1}$ and it decreases to 1\% at 5450 km s$^{-1}$.  
 \newline The integrated abundances of the most important elements in the photospheric ejecta, which is at a velocity above 4400 km s$^{-1}$, are M(Mg)=0.07 M$_{\sun}$, M(Fe)=0.03 M$_{\sun}$, M(O)=0.40 M$_{\sun}$, M(S)=0.058 M$_{\sun}$, M(Si)=0.09 M$_{\sun}$ and M($^{56}$Ni)=0.47 M$_{\sun}$. When the nebular phase spectra is available the $^{56}$Ni could increase to a total integrated abundance of 0.72 M$\sun$. The final total integrated abundances can be confirmed when late time spectra of SN 2014J are observed. Due to ground based telescopes not being able to observe the UV part of the spectra, the iron group element abundances of SN 2014J we have given here may show some 20\% uncertainty (of the values given), which nebular modelling will allow us to approve upon. 
\newline SN 2011fe and SN 2014J are photometrically similar. SN 2011fe has been used to determine the extinction of SN 2014J \citep{amanullah}, and it has also been modelled using the same MC radiative transfer code \citep{mazzali2013}. The total $^{56}$Ni abundance of SN 2011fe \citep{mazzali2013} and SN 2014J are very similar (0.4-0.7 M$_{\sun}$  and 0.47-0.72 M$_{\sun}$ respectively, the large range is due to not knowing abundance distribution in the nebula phase) Furthermore, the abundances in the photospheric region of the SN 2011fe ejecta are remarkably similar to those of SN 2014J. 
Changing the density profile in the models is not likely to qualitatively affect the abundance pattern in the regions of the ejecta explored by the spectra.

\section {Summary}
We have presented photometric and spectroscopic observation of the closest SN Ia in at least the last 28 years, SN 2014J. The observations were obtained with the LT and INT. We have presented SDSS \textit{g}$^{\prime}$\textit{r}$^{\prime}$\textit{i}$^{\prime}$ light curves and a spectral time series evolution over 12 epochs from --8d to +10d, relative to \textit{B}-band maximum. All of the spectra were calibrated in flux and atmospheric absorption lines were removed. The spectra show a very deep Si II 6355 \AA\ line and a Ca high velocity feature at $\sim$7900 \AA. We obtain a $\Delta$$m_{15}(B)$= 0.88$\pm$0.08 or 0.95$\pm$0.12 using the \textit{g}$^{\prime}$ and \textit{r}$^{\prime}$-band or \textit{r}$^{\prime}$-band respectively, and by fitting them through SIFTO. When correcting for reddening this produced values of  $\Delta$$m_{15}(B)$=0.98$\pm$0.08 or 1.05$\pm$0.12, respectively, using the SN 2011fe spectra, and  $\Delta$$m_{15}(B)$=0.95$\pm$0.08 or 1.02$\pm$0.12 for the Hsiao template. This result is consistent with those of \citet{Foley}. We obtain a V$_{max}$ of 10.66$\pm$0.02. 
\newline SN 2014J is a highly reddened SN Ia which does not follow the conventional Galactic reddening law (\textit{R}$_{V}$=3.1). We adopt the CCM law with a foreground galactic value of ${E(B-V)}$=0.05 (\textit{R}$_{V}$=3.1) and host galaxy extinction of ${E(B-V)}$=1.2 (\textit{R}$_{V}$=1.38). SN 2011fe and SN 2014J were found to have a comparable Ni masses, 0.4-0.7 M$\sun$  and 0.47-0.72 M$\sun$ respectively, when modelled using the same MC radiative transfer code. However, due to their different in uncorrected decline rate we could expect these values to change. This could be revealed with the use of the UV data and the nebular spectra.
\newline The spectra were modelled with the abundance tomography technique of \citet{Stehle2005}, simulating spectrum formation with a Monte-Carlo radiative transfer code. The density profile used for this was W7. In follow-up papers we will change the density structure in an attempt to refine the model.
\newline The spectra were modelled at 10 epochs, before and after \textit{B}-band maximum, inferring a best-fit abundance stratification. As one would expect, at higher velocities (12400 km s$^{-1}$) there is a large abundance of oxygen. As the photosphere recedes (8440 km s$^{-1}$) the IME elements dominate, Si and S. Then at low velocities the radioactive Ni dominates below 8000 km s$^{-1}$. This leads to the prediction that the characteristic line width of the iron emission line in the nebular spectra will be in the order of 8000 km s$^{-1}$ \citep{mazzali98}. 
\newline Synthetic spectra reproduce the spectral evolution of SN 2014J, and the final integrated abundances of SN 2014J are M(Mg)=0.07 M$_{\sun}$, M(Fe)=0.03 M$_{\sun}$, M(O)=0.40 M$_{\sun}$, M(S)=0.05 M$_{\sun}$, M(Si)=0.09 M$_{\sun}$ and M($^{56}$Ni)=0.47-0.72 M$_{\sun}$. Our results are consistent with the current understanding of SN Ia reddening and early time abundance distribution. The observation and modelling in this paper is of particular significance because of the close proximity of SN 2014J. Furthermore, SN 2014J is the typical example of a normal SN Ia, making our models a good basis for studying further objects.

\section*{Acknowledgments}
S.H. is funded from a Minerva ARCHES award of the German Ministry of Education and Research (BMBF). We have used data from the NASA / IPAC Extragalactic Database (NED, http://nedwww.ipac.caltech.edu, operated by the Jet Propulsion Laboratory, California Institute of Technology, under contract with the National Aeronautics and Space Administration). For data handling, we have made use of various software (as mentioned in the text) including \textsc{iraf}. \textsc{iraf}  Image Reduction and Analysis Facility (http://iraf.noao.edu) is an astronomical data reduction software distributed by the National Optical Astronomy Observatory (NOAO, operated by AURA, Inc., under contract with the National Science Foundation).

\bibliographystyle{apj}

\bsp

\end{document}